\documentclass[a4paper, 10pt, 3p, twocolumn, english]{elsarticle}
\usepackage{hyperref}

\journal{arXiv}

\begin{document}
\begin{frontmatter}

\title{Motion Planning for Triple-Axis Spectrometers}

\author[ill]{Tobias Weber \corref{cor}} \ead{tweber@ill.fr}
\address[ill]{Institut Laue-Langevin (ILL), 71 avenue des Martyrs, CS 20156, 38042 Grenoble Cedex 9, France}
\cortext[cor]{ORCID ID: \url{https://orcid.org/0000-0002-7230-1932}}

\begin{abstract}
We present the free and open source software \textit{TAS-Paths}, a novel system which calculates optimal,
collision-free paths for the movement of triple-axis spectrometers.
The software features an easy to use graphical user interface, but can also be scripted and used as a library.
It allows the user to plan and visualise the motion of the instrument before the experiment and can be used
during measurements to circumvent obstacles.
The instrument path is calculated in angular configuration space in order to keep a maximum angular distance from
any obstacle.
\end{abstract}

\end{frontmatter}

\section{Introduction}
Ever since their conception by Bertram Brockhouse, neutron triple-axis spectrometers (TAS)
have become an indispensable tool for research in solid-state physics.
They are regularly used in large-scale facilities around the world to probe nuclear or magnetic
excitations in crystals.

The three axis of a TAS instrument are the monochromator, the sample, and the analyser axis.
They are used for selecting the energy and momentum that is transferred from the neutrons to a sample crystal.
Modern instrument control software \cite{NICOS, Heer1997, Mutti2011} allows the user to drive the spectrometer using reciprocal crystal
coordinates which are internally translated by the software to rotations of the instrument axes and the sample crystal.
As the transformation from sample crystal coordinates to instrumental angles is not trivial \cite{Lumsden2005},
it is not immediately obvious to the user if a given coordinate leads to a collision of the
instrument with a shielding wall or any apparatus stationed in the instrument space.
The control software itself is usually only capable of catching the simplest pre-defined cases
of possible collisions, namely by allowing each motor to have soft limits in addition to the hard limits
imposed by end switches.
For details about the functioning of TAS please refer to the textbook by
Shirane, Shapiro, and Tranquada \cite{Shirane2002}.

Since the onset of the COVID-19 pandemic, both remote and autonomous experiments have come into focus.
Remote experimentation means that the users do not have to be at the instrument itself anymore;
they instead drive it via a remote control software.
Autonomous experiments go one step further and let the entire measurement be run by a machine learning system.
While remote control is already very common at neutron facilities, research into autonomous systems
based on Gaussian processes is currently being conducted at several facilities \cite{Noack2021, Parente2023}.

These new developments further disconnect the user from the instrument and exacerbate the problem of possible
collisions during scans or motor movement.
Even the machine learning software tools so far only consider the crystal coordinate space and are oblivious
to any details concerning the instrument, they specifically do not take into account how the coordinates are translated
into instrument coordinates and axis angles.

The present software is intended to bridge this gap and provides the means to calculate a safe path in
instrumental space from one set of either instrumental angles or crystal coordinates to another.
While collision detection and limited motion planning for triple-axis spectrometers has already been attempted in
the past \cite{Muehlbauer2006}, our algorithm goes beyond such a direct grid search and instead is based on
modern methods for analytical robot motion planning using Voronoi diagrams.

\section{Software Design}
The \textit{TAS-Paths} software is written in a modular way using modern \textit{C++20} and comprises a core library with
a scripting interface and a graphical user interface (GUI), which are each described in the following subsections.
\textit{TAS-Paths} makes heavy use of the \textit{Boost} library \cite{Boost} and utilises the same
mathematical library \cite{tlibs} that we employ in our triple-axis software suite \textit{Takin} \cite{Takin2021}.

\subsection{Core Library and Algorithm}
The basic algorithms of \textit{TAS-Paths}'s core module are realised as a \textit{C++} library.
The principal functionalities of the software comprise,
first, the building of a road map, i.e. a mesh of all possible paths on which the instrument can safely move,
and, second, picking the optimal path from a start to a target position from the mesh.
The approach is based on the retraction method (see \cite{Berg2008}, Ch. 7, pp. 163 and 304) of classical
motion planning for robots possessing full information about their surroundings.
It is named so because the robot retracts onto the road map, moves along its path mesh and detaches
in the vicinity of the goal.
A simplified version of the employed algorithm is presented in the following.

\begin{figure}[htb]
	\centering
	\includegraphics[width=0.8\columnwidth]{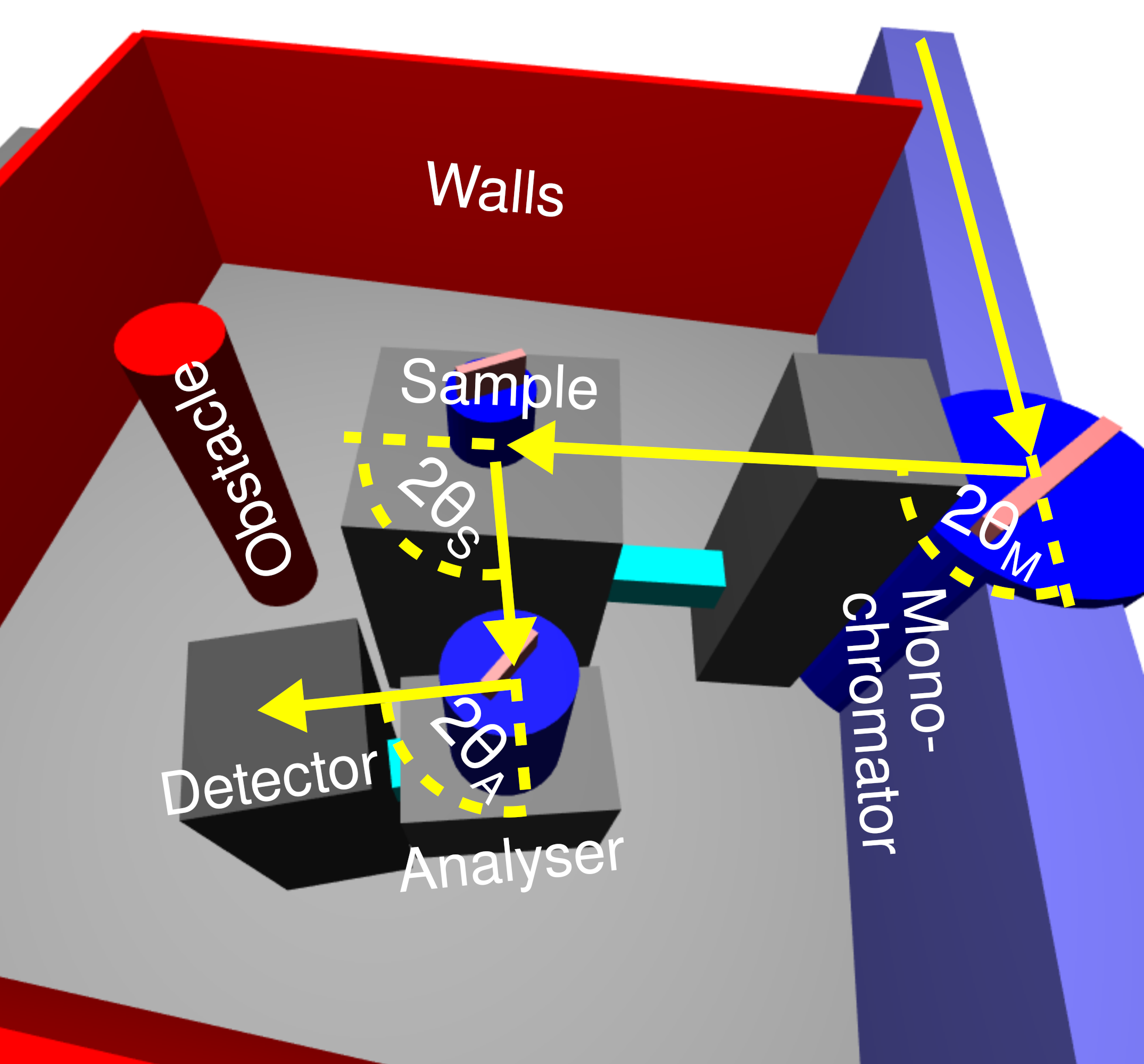}
	\caption{Instrumental space. During typical experiments, a triple-axis spectrometer uses two
		degrees of freedom of its axes, these are the sample axis angle $2\theta_S$ together with either
		the monochromator axis angle $2\theta_M$ or the analyser axis angle $2\theta_A$.
		This example shows a simplified model of the instrument \textit{MIRA} \cite{Georgii2018}
		and one obstacle.}
	\label{fig:path_real}
\end{figure}

\begin{figure}[htb]
	\centering
	\includegraphics[width=0.9\columnwidth]{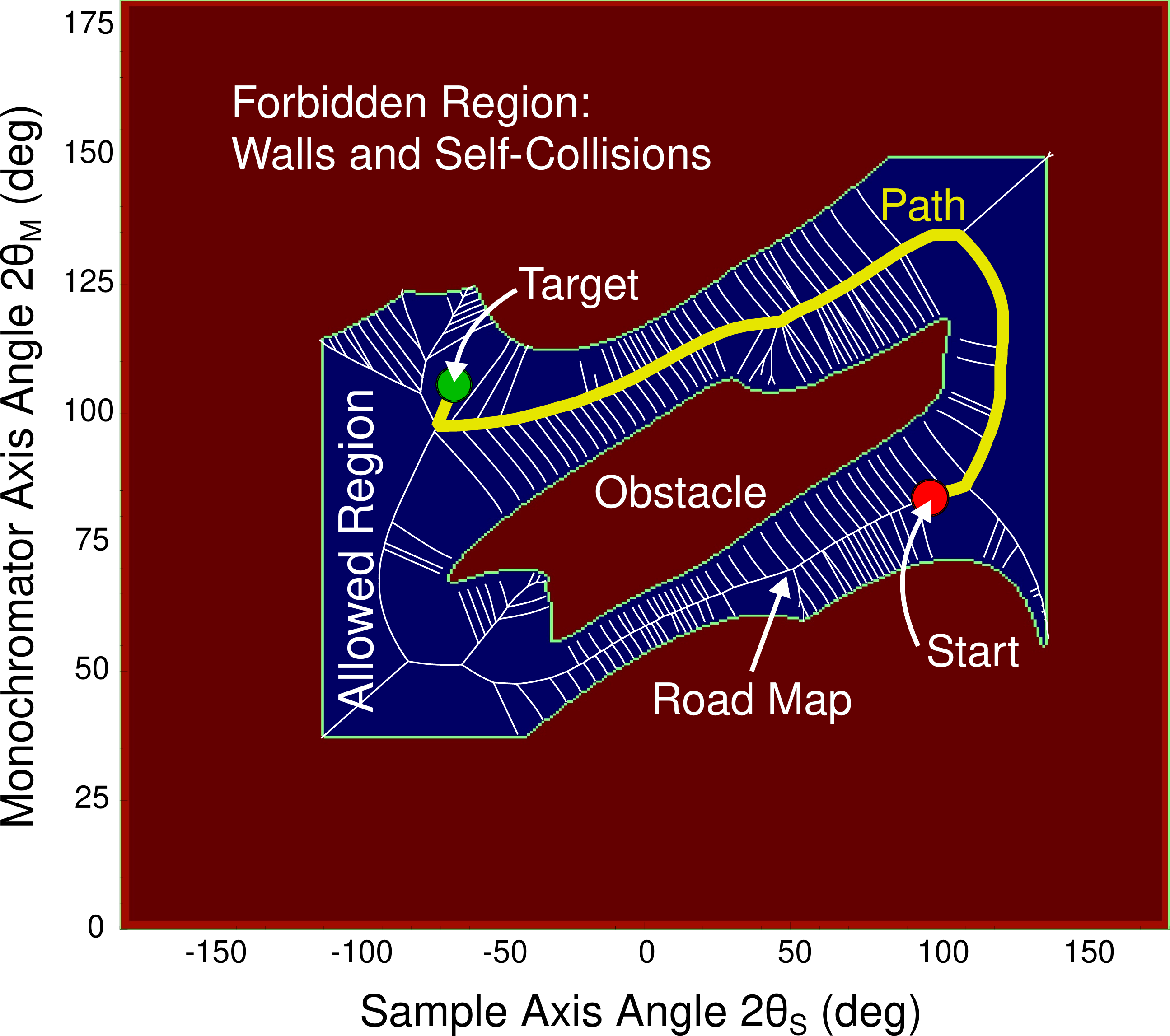}
	\caption{Angular configuration space spanned by the sample axis angle $2\theta_S$ and
		the monochromator axis angle $2\theta_M$. Forbidden regions leading to collisions are shown
		in red, while regions where the instrument can move are shown in blue.
		The calculated road map, i.e. the mesh of possible paths, is displayed as thin white lines.
		One concrete path from a start position (red circle) to a target position (green circle) is depicted
		as a thick yellow line.
		The figure corresponds to the instrumental space shown in Fig. \ref{fig:path_real}.}
	\label{fig:path_cfg}
\end{figure}

\paragraph{Path mesh calculation}
For the purpose of motion planning, a triple-axis instrument can be seen as a robotic arm possessing three
angular degrees of freedom (DOF), namely the monochromator, the sample, and the analyser axis,
as depicted in Fig. \ref{fig:path_real}. All DOF are constrained to the same two-dimensional plane.
Only two of the three DOF are used during an experiment, because energy selection can be done by
moving either the monochromator or the analyser axis, leaving the other axis at a fixed angle.
These two DOF, which are usually the monochromator and the sample axis angle for instruments operating in fixed
final energy mode, span a two-dimensional angular configuration space in which the position of the instrument
is represented as a single point. The configuration space corresponding to the instrument in
Fig. \ref{fig:path_real} is shown in Fig. \ref{fig:path_cfg}.
Please refer to \cite{Berg2008}, Ch. 13.1, pp. 284f for more information about configuration spaces in general.

It is in this two-dimensional angular configuration space that the road map as the mesh of possible
collision-free paths is constructed via a multi-step approach:
\begin{itemize}
\item In the first step, the software simulates instrumental positions on a grid in angular configuration
space and marks areas where the instrument would collide with obstacles in its path, with itself,
or areas where the axis motors are outside their angular limits.
This divides the configuration space into regions of allowed and disallowed positions, which
are shaded as blue and red areas in Fig. \ref{fig:path_cfg}, respectively.
\item The contours of these two regions' borders are reconstructed in the second step and approximated
by line segments $l_i$.
\item In the central step of the algorithm, the line segments $l_i$ serve as the sites for constructing
a Voronoi diagram $V$.
We calculate the mesh of possible instrument paths as the set of Voronoi bisectors
$B\left(l_i,\, l_j\right)$ of $V$ with \cite{Icking2001}:
\begin{equation}
	B\left(l_i,\, l_j\right) \ =\
		\left\{ \alpha \in C \left|\right.
			\left\Vert \alpha - l_i \right\Vert = \left\Vert \alpha - l_j \right\Vert \right\},
\end{equation}
where $C$ names the angular configuration space and $\left\Vert \cdot \right\Vert$ the angular norm.
Since $l_i$ and $l_j$ are any two line segments separating forbidden and allowed regions,
the bisectors give the loci of equal angular distance to the closest obstacles.
They are shown as the thin white lines in Fig. \ref{fig:path_cfg} and constitute the road map.
\end{itemize}

Using a Voronoi diagram ensures that the angular coordinates of each point on the path mesh
are at the farthest possible angular distance from any disallowed position.
Calculating the Voronoi diagram is done by one of two supported back end libraries.
The user can here choose between
\textit{Boost's Polygon} \cite{BoostPolygonVoronoi, Boost} and
\textit{CGAL's} \textit{Segment Delaunay Graph} \cite{Karavelas2006, CGAL} software.
For efficiency, the core library distributes the path mesh calculation onto multiple threads of execution.

The use of a line segment Voronoi diagram results from our quantisation of the boundaries of valid and invalid
configuration space into line segments.
In general, analytical terms, this is an approximation to the topological skeleton \cite{Golland2000}
of the configuration space.
Please refer to \cite{Berg2008}, Ch. 7, pp. 147-171 for more information on Voronoi diagrams and their generalisations.

\paragraph{Path calculation}
For performing calculations of paths, the Voronoi nodes and bisectors of the generated mesh are registered
as vertices and edges of a weighted graph data structure, respectively.
The angular distances between two Voronoi nodes serve as the weights of the graph edges,
where we also consider the speed of the axis motors for the calculation of the distance function.
A preliminary version of this kind of distance function has also served as cost function for our
complementary project concerning autonomous instrument control \cite{Noack2021}.
The shortest angular distance between a start and a target instrument angle is calculated
by employing Dijkstra's single-source shortest path algorithm (see \cite{Erickson2019}, pp. 273-297), which
finds the shortest path of a source vertex and all other vertices of a graph using the weights associated
with the edges.
In our example, the calculated shortest path is depicted as thick yellow line in Fig. \ref{fig:path_cfg}.

So far, the algorithm chooses the shortest path on the road map.
Nevertheless, there may still be alternative, safer paths available.
To calculate them, we optionally introduce additional weight factors depending on how far a position
is located from the next obstacle using the direct angular distance between them.
These weight factors effectively serve as a potential function pushing the optimal path away from
positions that are too close to forbidden regions.
The idea is similar to the classical potential field method of robot motion planning (see \cite{Berg2008}, p. 305),
where walls generate a repulsive potential to keep the robot at a distance,
but here with the addition of keeping the instrument on the road map.

The user can either directly supply the axis angles for the source and target position of the instrument or
alternatively specify Miller indices of the sample's crystal coordinate system.
The transformation from one into the other is done transparently by the software using the same
$UB$ matrix formalism \cite{Lumsden2005} that is also widely used by instrument control software,
e.g. \textit{NICOS} \cite{NICOS}, \textit{SICS} \cite{Heer1997}, or \textit{NOMAD} \cite{Mutti2011}.

\paragraph{External bindings}
The core \textit{C++} library features bindings to \textit{Python} \cite{Rossum2011} using
the \textit{SWIG} software \cite{SWIG} to generate a native interface.
This way calculations of instrument paths can be scripted and -- for instance -- be directly used in
\textit{Python}-based instrument control software like \textit{NICOS} \cite{NICOS}.
Reusable example scripts for both the fixed final and the fixed initial energy modes of TAS are provided
with the source code distribution.

\subsection{Graphical User Interface}
\begin{figure*}[htb]
	\centering
	\includegraphics[width=0.8\textwidth]{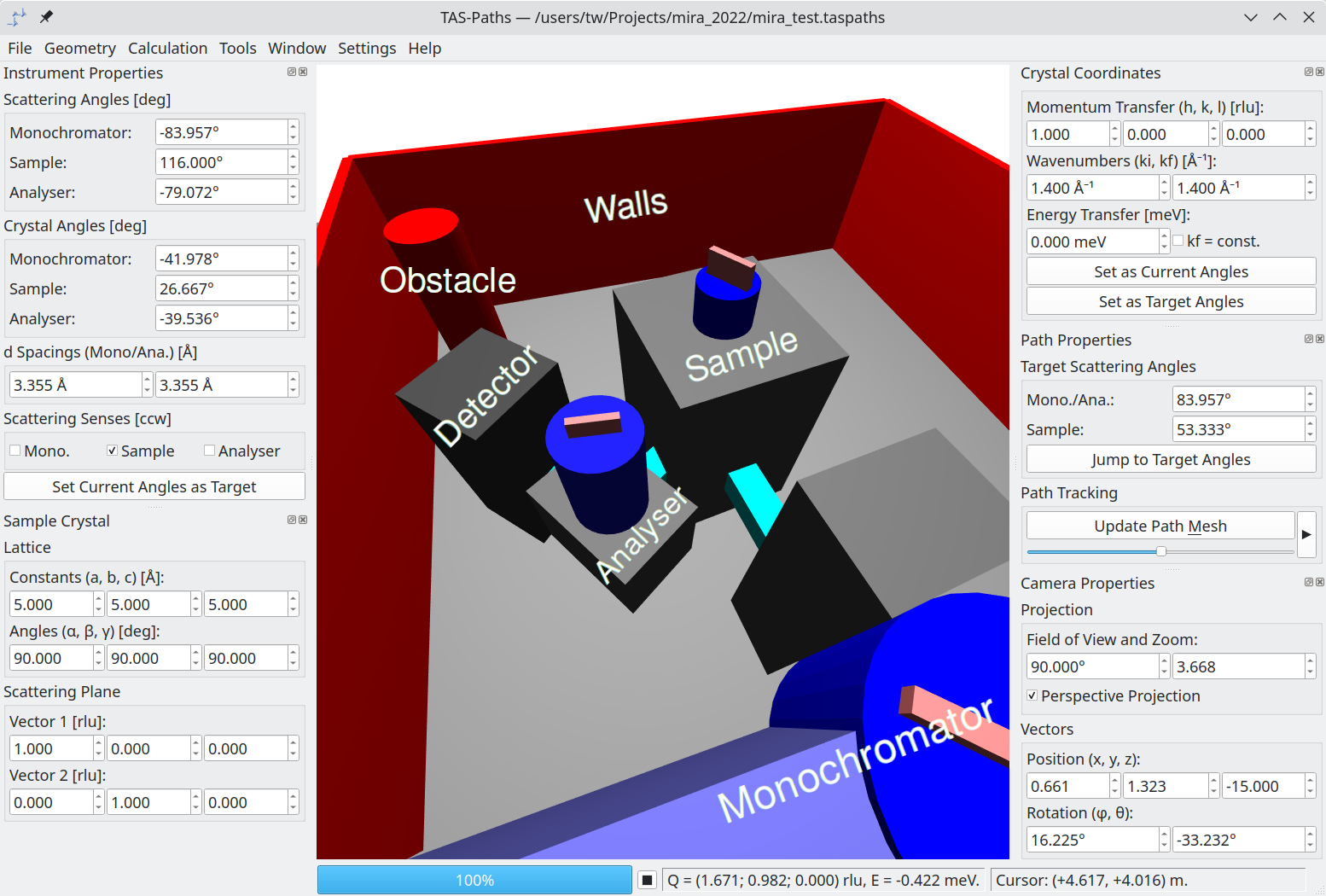}
	\caption{Main window. The GUI features a three-dimensional editor to easily add obstacles into the instrument space and move
		all components using mouse drag and drop.
		Instrumental motion along the calculated path is simulated and visualised.}
	\label{fig:taspaths}
\end{figure*}

The GUI is based on the \textit{Qt} library and its \textit{OpenGL} widgets \cite{Qt}.
It features a three-dimensional editor that lets the user place obstacles, move the instrument by dragging
and dropping the mouse and change its properties on-the-fly (Fig. \ref{fig:taspaths}).
For modeling the instrument and the obstacles we use geometric primitives.
These can be seen as bounding geometries to the actual, complicated instrumental details and as
such cannot lead to invalid positions.

The GUI uses the core library to calculate and simulate the path mesh and the specific path from the current
instrument position to a target position, which can also be chosen graphically.
The calculation and the instrumental movement in angular configuration space is visualised using
the \textit{QCustomPlot} library \cite{QCustomPlot}.
The calculated path can be exported to motor commands for several instrument control systems.

\section{Use Cases}
Apart from our main goal of including it into the autonomous instrument control project,
the software has multiple use cases of various degrees of complexity.
In the simplest case, it can be employed for calculating possible collision configurations when planning an experiment.
A more sophisticated scenario is to combine the software with an existing instrument control system
to automatically drive the instrument on a calculated path.
In the following we provide examples for these two cases.

\subsection{Collision calculation}
\begin{figure}[htb]
	\centering
	\includegraphics[width=0.95\columnwidth]{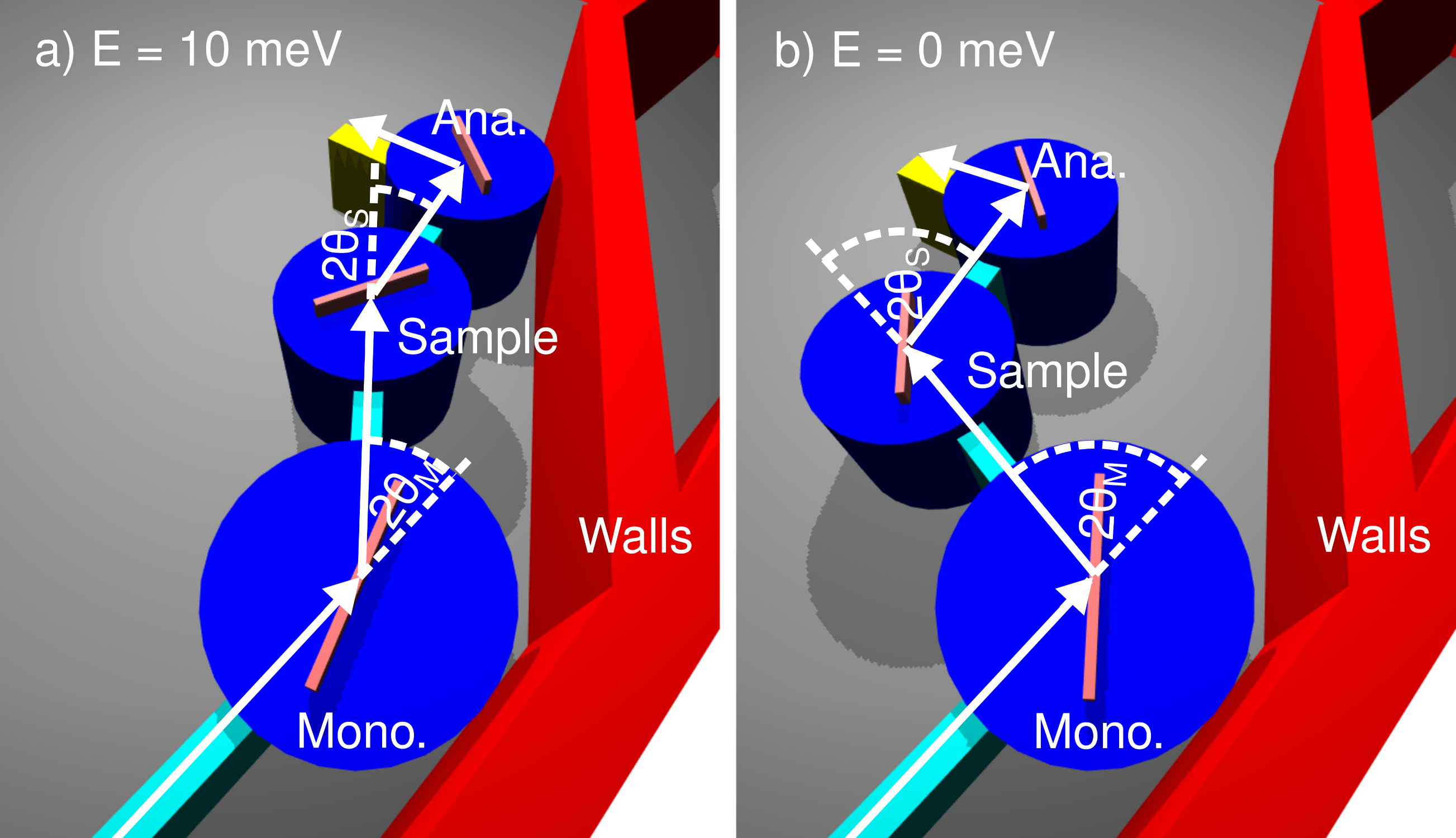}
	\caption{Without motion planning, moving an instrument that is at a position close
		to a wall (a), for example at the end of a scan, back to a neutral position (b)
		can lead to a crash into the wall if the monochromator axis moves slower than
		the sample axis.}
	\label{fig:near_wall}
\end{figure}

A collision that can easily occur during an experiment is due to the instrument's motors typically
moving at different speeds.
This can lead to a situation where a drive command issued to an instrument that is
already close to a wall can crash the instrument into the wall even if both the source and the
target positions by themselves would be safe.
Such a situation is shown in Fig. \ref{fig:near_wall}, it is modeled after scan positions
we encountered during an experiment at the instrument \textit{Thales} \cite{Boehm2015}.
At \textit{Thales}, the monochromator axis moves considerably more slowly than the sample axis.
Manually driving back the instrument from position (a) in Fig. \ref{fig:near_wall},
which was at the end of a scan, to its neutral position (b) would therefore have led
to collisions with the wall, because the control software usually moves all axes at the same time.

In such a scenario, the \textit{TAS-Paths} core library might be linked to the instrument control
software and used as a monitoring system warning of imminent collisions when the control software
issues a drive command.

\subsection{Path tracking}
Apart from mere monitoring of the positions driven by the instrument control system, \textit{TAS-Paths}
can actively calculate the paths to be used by the control software.
We successfully performed first tests at the instrument \textit{MIRA} \cite{Georgii2018}
using the \textit{NICOS} \cite{NICOS} control system, where we let the software drive the instrument around an obstacle.
For our test, a trash can served as a placeholder for the obstacle.
In a real experiment, this could typically be a helium or nitrogen can or devices dedicated to
sample environment which usually tend to be placed close to the instrument arms.
A video of the experiment is available on-line at \url{https://youtu.be/F0SAQp00he4}.
Note that due to slow instrument movement the video has been sped up multifold for easier viewing.

The results of the test show that in order to be compatible with \textit{TAS-Paths}, the instrument
control system has to be adapted for a different kind of motor control.
In normal operation, the control system would drive all motors simultaneously, correct their
positions within less than a tenth of a degree of precision and switch off the air cushions after each point.
For \textit{TAS-Paths}, the movement has to be continuous, keeping the air cushions on.
A correction after each control point within the degree range is usually sufficient.

\section{Conclusion}
We presented the novel collision detection and motion planning system \textit{TAS-Paths}.
The software finds the instrumental path which has the farthest angular distance from any
obstacles that may be in the experiment space of a triple-axis spectrometer.

As there is currently no automated safety system in place during an autonomous experiment
that would prevent the instrument from collisions, the software will be included in our larger
autonomous experimentation project for which we have performed first tests \cite{Noack2021}
and which will be available for normal user operation in the future.
This is possible since \textit{TAS-Paths} is complementary to our current autonomous control system
in every respect:
\textit{TAS-Paths} calculates its positions analytically and is only concerned about the instrumental
side of an experiment, while the autonomous control software uses statistical methods and
heuristics \cite{Noack2023} for position selection and concerns only the physics side of the experiment.

Aside from its inclusion in autonomous experiments, the \textit{TAS-Paths} software provides
a means for the user to simulate and visualise instrument movement during experiment planning.
This will help prevent dangerous situations before commencing any measurements.
The software has been successfully tested for typical situations at instruments as well as
for moving a spectrometer around obstacles.

Future developments might include the automatic determination of obstacles in instrument
space using RGBZ cameras and the usage of external, full CAD models of
instruments instead of simplified geometries.

\section*{Acknowledgements}
First and foremost I am indebted to Dr. Lihong Ma and Prof. Dr. Christian Icking for supervising this project
as part of a thesis work \cite{ThesisWeber} and for many discussions about it and computational geometry in general.
I wish to thank Ga\'etan Fine for his work on integrating \textit{TAS-Paths} with the \textit{NOMAD}
instrument control system \cite{Mutti2011} and for beta testing.
Further thanks go to Robert Georgii for his support with the \textit{MIRA} \cite{Georgii2018} commissioning test,
to Martin B\"ohm and Paul Steffens for discussions, and to Paolo Mutti and St\'ephane Rols for supporting the project.
Thanks to Alain Filhol for his help getting the software onto the \textit{App Store}.

\section*{Source Code}
\textit{TAS-Paths's} unified release repository is located at \url{https://github.com/ILLGrenoble/taspaths},
its development repository can be found at \url{https://code.ill.fr/scientific-software/takin/paths}.
The software's DOI is \href{https://doi.org/10.5281/zenodo.4625649}{10.5281/zenodo.4625649}.


\begin{thebibliography}{10}
\expandafter\ifx\csname url\endcsname\relax
  \def\url#1{\texttt{#1}}\fi
\expandafter\ifx\csname urlprefix\endcsname\relax\def\urlprefix{URL }\fi
\expandafter\ifx\csname href\endcsname\relax
  \def\href#1#2{#2} \def\path#1{#1}\fi

\bibitem{NICOS}
G.~Brandl, E.~Faulhaber, C.~Felder, J.~Kr\"uger, A.~Lenz, B.~Pedersen,
  \href{https://nicos-controls.org}{{NICOS, the Networked Instrument Control
  System}} (2010 - 2023).
\newline\urlprefix\url{https://nicos-controls.org}

\bibitem{Heer1997}
H.~Heer, M.~K\"onnecke, D.~Maden, {The SINQ instrument control software
  system}, Physica B: Condensed Matter 241-243 (1997) 124--126, proceedings of
  the International Conference on Neutron Scattering.
\newblock \href {https://doi.org/10.1016/S0921-4526(97)00528-0}
  {\path{doi:10.1016/S0921-4526(97)00528-0}}.

\bibitem{Mutti2011}
P.~Mutti, F.~C{\'e}cillon, A.~Elaazzouzi, Y.~L. Goc, J.~Locatelli, H.~Ortiz,
  J.~Ratel, {NOMAD -- more than a simple sequencer}, in: Proceedings of
  ICALEPCS 2011, Grenoble, France, 2011, pp. 808 -- 811.

\bibitem{Lumsden2005}
M.~D. Lumsden, J.~L. Robertson, M.~Yethiraj, {UB matrix implementation for
  inelastic neutron scattering experiments}, Journal of Applied Crystallography
  38~(3) (2005) 405--411.
\newblock \href {https://doi.org/10.1107/S0021889805004875}
  {\path{doi:10.1107/S0021889805004875}}.

\bibitem{Shirane2002}
G.~Shirane, S.~M. Shapiro, J.~M. Tranquada, {Neutron Scattering with a
  Triple-Axis Spectrometer: Basic Techniques}, Cambridge University Press,
  Cambridge, UK, 2002.

\bibitem{Noack2021}
M.~M. Noack, P.~H. Zwart, D.~M. Ushizima, M.~Fukuto, K.~G. Yager, K.~C. Elbert,
  C.~B. Murray, A.~Stein, G.~S. Doerk, E.~H.~R. Tsai, R.~Li, G.~Freychet,
  M.~Zhernenkov, H.-Y.~N. Holman, S.~Lee, L.~Chen, E.~Rotenberg, T.~Weber,
  Y.~L. Goc, M.~B\"ohm, P.~Steffens, P.~Mutti, J.~A. Sethian, {Gaussian
  processes for autonomous data acquisition at large-scale synchrotron and
  neutron facilities}, Nature Review Physics 3 (2021) 685--697.
\newblock \href {https://doi.org/10.1038/s42254-021-00345-y}
  {\path{doi:10.1038/s42254-021-00345-y}}.

\bibitem{Parente2023}
M.~T. Parente, G.~Brandl, C.~Franz, U.~Stuhr, M.~Ganeva, A.~Schneidewind,
  \href{https://arxiv.org/abs/2209.00980}{{AI-assisted neutron spectroscopy
  using active learning with log-Gaussian processes}}, preprint available on
  arXiv: 2209.00980 (2023).
\newblock \href {http://arxiv.org/abs/2209.00980} {\path{arXiv:2209.00980}}.
\newline\urlprefix\url{https://arxiv.org/abs/2209.00980}

\bibitem{Muehlbauer2006}
Q.~M\"uhlbauer, K.~Hradil, {Monitoring and Preventing Collisions for a Triple
  Axis Spectrometer}, in: {Proceedings of the 2006 IEEE International
  Conference on Control Applications, Munich, Germany, October 4-6, 2006},
  2006, pp. 1795--1800.
\newblock \href {https://doi.org/10.1109/CACSD-CCA-ISIC.2006.4776913}
  {\path{doi:10.1109/CACSD-CCA-ISIC.2006.4776913}}.

\bibitem{Boost}
{Various Authors}, \href{https://www.boost.org}{{Boost C++ Libraries}}
  (2004-2023).
\newline\urlprefix\url{https://www.boost.org}

\bibitem{tlibs}
T.~Weber, {tlibs -- a physical-mathematical C++ template library} (2012-2023).
\newblock \href {https://doi.org/10.5281/zenodo.5717779}
  {\path{doi:10.5281/zenodo.5717779}}.

\bibitem{Takin2021}
T.~Weber, {Update 2.0 to ``Takin: An open-source software for experiment
  planning, visualisation, and data analysis'', (PII: S2352711016300152)},
  SoftwareX 14 (2021) 100667.
\newblock \href {https://doi.org/10.1016/j.softx.2021.100667}
  {\path{doi:10.1016/j.softx.2021.100667}}.

\bibitem{Berg2008}
M.~de~Berg, O.~Cheong, M.~van Kreveld, M.~Overmars, {Computational Geometry,
  Algorithms and Applications, Third Edition}, Springer-Verlag Berlin
  Heidelberg, Germany, 2008.
\newblock \href {https://doi.org/10.1007/978-3-540-77974-2}
  {\path{doi:10.1007/978-3-540-77974-2}}.

\bibitem{Georgii2018}
R.~Georgii, T.~Weber, G.~Brandl, M.~Skoulatos, M.~Janoschek, S.~M\"uhlbauer,
  C.~Pfleiderer, P.~B\"oni, {The multi-purpose three-axis spectrometer (TAS)
  MIRA at FRM II}, Nuclear Instruments and Methods in Physics Research Section
  A: Accelerators, Spectrometers, Detectors and Associated Equipment 881 (2018)
  60--64.
\newblock \href {https://doi.org/10.1016/j.nima.2017.09.063}
  {\path{doi:10.1016/j.nima.2017.09.063}}.

\bibitem{Icking2001}
C.~Icking, R.~Klein, L.~Ma, S.~Nickel, A.~Wei\ss{}ler, {On bisectors for
  different distance functions}, Discrete Applied Mathematics 109~(1) (2001)
  139--161, 14th European Workshop on Computational Geometry.
\newblock \href {https://doi.org/10.1016/S0166-218X(00)00238-9}
  {\path{doi:10.1016/S0166-218X(00)00238-9}}.

\bibitem{BoostPolygonVoronoi}
A.~Sydorchuk,
  \href{https://www.boost.org/doc/libs/1_76_0/libs/polygon/doc/voronoi_main.htm}{{Boost
  Polygon Voronoi extensions, part of the Boost C++ library \cite{Boost}}}
  (2010-2013).
\newline\urlprefix\url{https://www.boost.org/doc/libs/1_76_0/libs/polygon/doc/voronoi_main.htm}

\bibitem{Karavelas2006}
M.~I. Karavelas,
  \href{http://users.math.uoc.gr/~mkaravel/files/papers/vda-ewcg06.pdf}{{Voronoi
  diagrams in CGAL \cite{CGAL}}}, in: I.~Z. Emiris, M.~I. Karavelas, L.~Palios
  (Eds.), {Proceedings of the 22nd European Workshop on Computational Geometry
  (EWCG'06)}, Delphi, Greece, 2006, pp. 229--232.
\newline\urlprefix\url{http://users.math.uoc.gr/~mkaravel/files/papers/vda-ewcg06.pdf}

\bibitem{CGAL}
{Various Authors}, \href{https://www.cgal.org}{{Computational Geometry
  Algorithms Library (CGAL) for C++}} (1995-2023).
\newline\urlprefix\url{https://www.cgal.org}

\bibitem{Golland2000}
P.~Golland, W.~Eric, L.~Grimson, Fixed topology skeletons, in: {Proceedings
  IEEE Conference on Computer Vision and Pattern Recognition. CVPR 2000 (Cat.
  No.PR00662)}, Vol.~1, 2000, pp. 10--17.
\newblock \href {https://doi.org/10.1109/CVPR.2000.855792}
  {\path{doi:10.1109/CVPR.2000.855792}}.

\bibitem{Erickson2019}
J.~Erickson,
  \href{http://jeffe.cs.illinois.edu/teaching/algorithms/}{{Algorithms, 1st
  edition}}, self-published, 2019.
\newline\urlprefix\url{http://jeffe.cs.illinois.edu/teaching/algorithms/}

\bibitem{Rossum2011}
G.~van Rossum, J.~Fred L.~Drake, {Python Language Reference Manual}, Network
  Theory Ltd., 2011.

\bibitem{SWIG}
{The SWIG Developers}, \href{http://www.swig.org}{{Simplified Wrapper and
  Interface Generator (SWIG)}} (1995-2023).
\newline\urlprefix\url{http://www.swig.org}

\bibitem{Qt}
{The Qt Company}, \href{https://www.qt.io}{{Qt}} (2023).
\newline\urlprefix\url{https://www.qt.io}

\bibitem{QCustomPlot}
E.~Eichhammer, \href{https://www.qcustomplot.com}{{QCustomPlot}} (2011 - 2023).
\newline\urlprefix\url{https://www.qcustomplot.com}

\bibitem{Boehm2015}
M.~B\"ohm, P.~Steffens, J.~Kulda, M.~Klicpera, S.~Roux, P.~Courtois,
  P.~Svoboda, J.~Saroun, V.~Sechovsky, {ThALES -- Three Axis Low Energy
  Spectroscopy for highly correlated electron systems}, Neutron News 26~(3)
  (2015) 18--21.
\newblock \href {https://doi.org/10.1080/10448632.2015.1057050}
  {\path{doi:10.1080/10448632.2015.1057050}}.

\bibitem{Noack2023}
M.~M. Noack, K.~G. Reyes, {Mathematical nuances of Gaussian process-driven
  autonomous experimentation}, MRS Bulletin 48 (2023) 1--11.
\newblock \href {https://doi.org/10.1557/s43577-023-00478-8}
  {\path{doi:10.1557/s43577-023-00478-8}}.

\bibitem{ThesisWeber}
T.~Weber, {Pathfinding for Triple-Axis Spectrometers}, {FernUniversit\"at in
  Hagen (FUH), Fakult\"at f\"ur Mathematik und Informatik, Lehrgebiet
  Kooperative Systeme, Hagen, Germany} (2021).
\end{thebibliography}
\end{document}